\documentstyle[aps,prd,twocolumn]{revtex}

\begin{document}
\title{Singularities in axially symmetric solutions of Einstein-Yang Mills and
related theories}
\author{Ludger Hannibal}
\address{Fachbereich Physik, Carl von Ossietzky Universit\"{a}t Oldenburg, D-26111\\
Oldenburg, Germany}
\date{March 8th, 1999}
\maketitle
\pacs{04.20.Jb}

\begin{abstract}
We show that the solutions of SU(2) Yang-Mills-dilation and
Einstein-Yang-Mills-dilaton theories described in a sequence of papers by
Kleihaus and Kunz are not regular in the gauge field part.
\end{abstract}

Since the pioneering papers by Bogomol'nyi \cite{bogomolnyi1976} and Witten 
\cite{witten1977} on exact solutions of classical Yang-Mills theory many
studies in this field have been carried out. In recent time the
gravitational self-interaction of solitons has been of interest \cite
{volkov1999}. Not only spherically symmetric solutions of theories including
Yang-Mills fields, with or without gravitation, Higgs, or dilaton fields,
have been constructed, but also solutions with axial symmetry \cite
{rebbi1980}. These latter type of solitons was studied by Kleihaus and Kunz
in the context of Yang-Mills-dilaton theory \cite{kleihaus1997a} and
Einstein-Yang-Mills-dilaton theory \cite
{kleihaus1997b,kleihaus1997c,kleihaus1998a,kleihaus1998b}, mainly by
numerical construction. The relevance of solutions containing black holes is
that they should provide counterexamples to the no-hair conjecture for such
theories. This conjecture is the analog to the no-hair theorems in
Einstein-Maxwell and related theories \cite{heusler1996}.

In this communication we would like to point out that the gauge field part
of the solutions described in Refs. \cite
{kleihaus1997a,kleihaus1997b,kleihaus1997c,kleihaus1998a,kleihaus1998b} are
not regular on the symmetry axis, in contradiction to what is claimed. Thus
the solutions are similar to singular monopole solutions and thus have to be
interpreted carefully \cite{wentzel1966,arafune1975,wu1975}. We explicitly
derive conditions for the regularity, and show that a recent attempt by
Kleihaus \cite{kleihaus1999a} to relate the non-regular solution to regular
ones via a (non-regular) gauge transformations like it is done in the case
of monopoles falls short of its aim, so that one must ask whether the
solutions described exist in any well-defined way.

Non-regular gauge fields are encountered within Dirac's theory of monopoles.
In general, solutions of \ Maxwell's equations for the electromagnetic
four-potential must be twice continuous differentiable in order to be
acceptable from the mathematical point of view. From the Aharonov-Bohm
effect we know that continuity of \ the potential is essential. For the
derivation of the field equations from an action one uses partial
integration, thus the solutions must be differentiable. In general,
regularity means that all solutions are at least twice differentiable in the
sense of distributions, allowing for delta-like sources. Vacuum solution
will in general be locally analytic. In Dirac's theory of monopoles there
exists no globally defined regular gauge field, only the electromagnetic
field tensor is regular. One can find a gauge transformation in the vicinity
of the singularity on the symmetry axis that gives a locally regular gauge
field, and the compatibility condition in the overlap region leads to the
quantization of the monopole \cite{wentzel1966,arafune1975}. In the case of
non-abelian gauge theories the situation is more complicated , and for
overlapping solutions connected by gauge transformations compatibility
conditions are required \cite{wu1975}. Another problem is that the gauge
potential explicitly enter in the field equations for the gauge field
tensor, because in contrast to the abelian case the gauge field tensor is
only gauge covariant, not invariant. Thus, at singular points of the gauge
potential the field equations are not well defined, even if the field tensor
is. In the case of spherically symmetric solutions there will in general be
only a singularity at the origin, the coordinate singularity on the $z$-axis
plays no role due to the spherical symmetry. Witten \cite{witten1977}
explicitly showed how such a singularity can be removed by a gauge
transformation.

In the case of axially symmetric solutions the situation becomes more
complicated, since a coordinate singularity at $\theta =0,\pi $ is
introduced, but cannot be removed by a symmetry argument. Together with
axial symmetry Rebbi and Rossi \cite{rebbi1980} introduced a winding number $%
n$ into their ansatz for the gauge field, using unit vectors in the Lie
algebra $su(2)$ of $SU(2)\,$that correspond to cylindrical coordinates $%
\left( \rho ,\phi ,z\right) $. Writing these as 
\begin{eqnarray}
u_{1} &=&\tau _{\rho }^{n}=\tau _{1}\cos n\phi +\tau _{2}\sin n\phi 
\nonumber \\
u_{2} &=&\tau _{\phi }^{n}=-\tau _{1}\sin n\phi +\tau _{2}\cos n\phi
\label{1} \\
u_{3} &=&\tau _{z}^{n}=\tau _{3}  \nonumber
\end{eqnarray}
where $\tau _{i},i=1,2,3$ are the Pauli matrices, the gauge field attains
the form 
\begin{equation}
A_{\mu }=u_{i}W_{\mu }^{i}  \label{3}
\end{equation}
with twelve independent gauge fields $W_{\mu }^{i},i=1,2,3,\mu =0,1,2,3$. In
spherical coordinates $\left( r,\theta ,\phi \right) $ the unit vector of $%
su(2)$ will be 
\begin{eqnarray}
\tau _{r}^{n} &=&\tau _{1}\sin \theta \cos n\phi +\tau _{2}\sin \theta \sin
n\phi +\tau _{3}\cos \theta ,  \nonumber \\
\tau _{\theta }^{n} &=&\tau _{1}\cos \theta \cos n\phi +\tau _{2}\cos \theta
\sin n\phi -\tau _{3}\sin \theta ,  \label{2} \\
\tau _{\phi }^{n} &=&-\tau _{1}\sin n\phi +\tau _{2}\cos n\phi  \nonumber
\end{eqnarray}
The gauge potentials $A_{\mu }$ are not regular unless the functions $W_{\mu
}^{i}$ behave appropriately on the $z$-axis, since the functions $\sin n\phi
,\cos n\phi $, when considered as functions on $S^{2}$, are neither
continuous nor differentiable at the poles $\theta =0,\pi $. The functions $%
\sin n\phi $ and $\cos n\phi $ can be expanded into polynomials of degree $%
\left| n\right| $ in the variables $\sin \phi =y/\rho $ and $\cos \phi
=x/\rho $, with $\rho =\sqrt{x^{2}+y^{2}}$; all terms of these polynomials
have denominators that are even powers of $\rho $ for $n$ even, and odd
powers of $\rho $ for $n$ odd. Hence $\rho ^{\left| n\right| }\sin n\phi $
and $\rho ^{\left| n\right| }\cos n\phi $ are regular functions except for $%
r=0$, and may further be multiplied by functions of the variables $\rho
^{2}\,$and $z$ without loosing their regularity. This was explicitly taken
into account by Rebbi and Rossi \cite{rebbi1980}. In spherical coordinates
functions of the type 
\begin{equation}
f\left( r,\theta ^{2}\right) \sin ^{\left| n\right| }\theta \sin n\phi
+g\left( r,\theta ^{2}\right) \sin ^{\left| n\right| }\theta \cos n\phi
\label{4}
\end{equation}
with regular functions $f$ and $g$ are regular along the axis $\theta =0$,
possibly except at $r=0$. The well defined $\theta $-parity of these
functions is easily understood. On any great circle through the pole at $%
\theta =0$, with $\phi =\phi _{0}$ to the right of the pole, $\phi =\phi
_{0}\pm \pi $ to the left, and coordinate 
\begin{equation}
\alpha =\left\{ 
\begin{array}{c}
\theta \text{ for }\alpha >0,\phi =\phi _{0} \\ 
-\theta \text{ for }\alpha <0,\phi =\phi _{0}\pm \pi
\end{array}
\right.  \label{5}
\end{equation}
we have for any function $F\left( \theta \right) $ (defined only for $\theta
>0$) 
\begin{eqnarray}
\lim_{\alpha \nearrow 0}\frac{\partial ^{k}}{\partial \alpha ^{k}}F(\theta
&=&-\alpha )\sin n\left( \phi _{0}\pm \pi \right)  \nonumber \\
\qquad &=&\left( -1\right) ^{n+k}\lim_{\alpha \searrow 0}\frac{\partial ^{k}%
}{\partial \alpha ^{k}}F\left( \theta =\alpha \right) \sin n\phi _{0}.
\end{eqnarray}
Therefore it is necessary that $F\left( \theta \right) $ is an even function
in $\theta $ for $n$ even, an odd function for $n$ odd in order that $%
F\left( \theta \right) \sin n\phi $ can be continuous $\left( k=0\right) $
or that the derivatives exist $\left( k=1,2,...\right) $. Together with the
multiplication by the factor $\sin ^{\left| n\right| }\theta $ we obtain (%
\ref{4}) as a sufficient form for functions to be analytical at $\theta =0$.

Kleihaus and Kunz use spherical coordinates $\left( r,\theta ,\phi \right) $%
. Their ansatz for the $SU(2)$ gauge field is \cite{kleihaus1998a}: 
\begin{equation}
A_{\mu }dx^{\mu }=\frac{1}{2gr}\left\{ 
\begin{array}{l}
\displaystyle{\tau _{\phi }^{n}\left[ H_{1}dr+(1-H_{2})rd\theta \right]}  \\ 
\displaystyle{
-n\left[ \tau _{r}^{n}H_{3}+\tau _{\theta }^{n}\left( 1-H_{4}\right) \right]
r\sin \theta d\phi }
\end{array}
\right\}   \label{7}
\end{equation}
This gauge potential is singular at $x=y=0$ unless appropriate conditions
are imposed on the functions $H_{i}$, $i=1,2,3,4$ that depend only on $r$
and $\theta $. In order to analyze this ansatz for regularity, we decompose
the gauge potential into its Cartesian components, 
\begin{equation}
A_{i}dx^{i}=\frac{1}{2gr}\sum_{j=1}^{3}\tau ^{j}\left(
w_{1}^{j}dx+w_{2}^{j}dy+w_{3}^{j}dz\right) ,
\end{equation}
with the nine functions $w_{i}^{j}$ given by 
\begin{eqnarray}
w_{1}^{1} &=&-F_{1}\sin n\phi \cos \phi +nF_{3}\cos n\phi \sin \phi  
\nonumber \\
w_{1}^{2} &=&F_{1}\cos n\phi \cos \phi +nF_{3}\sin n\phi \sin \phi  
\nonumber \\
w_{1}^{3} &=&nF_{4}\sin \phi   \nonumber \\
w_{2}^{1} &=&-F_{1}\sin n\phi \sin \phi -nF_{3}\cos n\phi \cos \phi  
\nonumber \\
w_{2}^{2} &=&F_{1}\cos n\phi \sin \phi -nF_{3}\sin n\phi \cos \phi  
\nonumber \\
w_{2}^{3} &=&-nF_{4}\cos \phi   \nonumber \\
w_{3}^{1} &=&-F_{2}\sin n\phi   \nonumber \\
w_{3}^{2} &=&F_{2}\cos n\phi   \nonumber \\
w_{3}^{3} &=&0.
\end{eqnarray}
The functions $F_{1}$ to $F_{4}$ are defined by 
\begin{eqnarray}
F_{1} &=&H_{1}\sin \theta +(1-H_{2})\cos \theta   \nonumber \\
F_{2} &=&H_{1}\cos \theta -(1-H_{2})\sin \theta   \nonumber \\
F_{3} &=&H_{3}\sin \theta +(1-H_{4})\cos \theta \   \nonumber \\
F_{4} &=&H_{3}\cos \theta -(1-H_{4})\sin \theta .  \label{9}
\end{eqnarray}
We see that a sufficient condition that the gauge potential can be
analytically extended to $\theta =0$ for $r>0\,$is 
\begin{eqnarray}
F_{1}\left( r,\theta \right)  &=&\tilde{F}_{1}\left( r,\theta ^{2}\right)
\sin ^{\left| n\right| +1}\theta   \label{10a} \\
F_{2}\left( r,\theta \right)  &=&\tilde{F}_{2}\left( r,\theta ^{2}\right)
\sin ^{\left| n\right| }\theta   \label{10b} \\
F_{3}\left( r,\theta \right)  &=&\tilde{F}_{3}\left( r,\theta ^{2}\right)
\sin ^{\left| n\right| +1}\theta   \label{10c} \\
F_{4}\left( r,\theta \right)  &=&\tilde{F}_{4}\left( r,\theta ^{2}\right)
\sin \theta   \label{10d}
\end{eqnarray}
with analytic functions $\tilde{F}_{1}$ to $\tilde{F}_{4}$. If regularity is
sought only up to a certain number of derivatives, then accordingly less
constraints on the odd orders in $\theta \,$are needed. Now a more subtle
point is that these conditions guarantee regularity, but are not necessary.
In the special case $n=1$ with spherical symmetry and $%
H_{1}=H_{3}=0,H_{2}=H_{4}=w(r)$ we see that $F_{1}$ and $F_{3}$ do not
respect (\ref{10a}) and (\ref{10c}), respectively. The reason is that in the
case $F_{1}=F_{3}$ the relation $\cos ^{2}\phi +\sin ^{2}\phi =1$ leads to a
regular function where in general $F_{1}\cos ^{2}\phi +F_{3}\sin ^{2}\phi $
would not have been regular. It is not difficult to see that for arbitrary $n
$ this will happen if 
\begin{equation}
F_{1}=nF_{3}
\end{equation}
so that we can do with two powers less of $\sin \theta $ in (\ref{10a}) and (%
\ref{10c}) if the functions agree in the lowest order. Hence an ansatz that
is analytical and also respects all symmetry requirements at $\theta =0,\pi
/2,\pi $ is given by 
\begin{eqnarray}
F_{1}\left( r,\theta \right)  &=&\left[ nf\left( r\right) +\tilde{F}%
_{1}\left( r,\sin ^{2}\theta \right) \sin ^{2}\theta \right] \cos \theta
\sin ^{\left| n\right| -1}\theta   \label{10aa} \\
F_{2}\left( r,\theta \right)  &=&\tilde{F}_{2}\left( r,\sin ^{2}\theta
\right) \sin ^{\left| n\right| }\theta   \label{10ab} \\
F_{3}\left( r,\theta \right)  &=&\left[ f\left( r\right) +\tilde{F}%
_{3}\left( r,\sin ^{2}\theta \right) \sin ^{2}\theta \right] \cos \theta
\sin ^{\left| n\right| -1}\theta   \label{10ac} \\
F_{4}\left( r,\theta \right)  &=&\tilde{F}_{4}\left( r,\sin ^{2}\theta
\right) \sin \theta   \label{10ad}
\end{eqnarray}
For the function $H_{i}$ this implies 
\begin{eqnarray}
H_{1} &=&-F_{1}\sin \theta +F_{2}\cos \theta   \nonumber \\
&=&\left[ -nf\left( r\right) -\tilde{F}_{1}\sin ^{2}\theta +\tilde{F}_{2}%
\right] \cos \theta \sin ^{\left| n\right| }\theta   \nonumber \\
(1-H_{2}) &=&F_{1}\cos \theta +F_{2}\sin \theta   \nonumber \\
&=&\left[ \left( nf\left( r\right) +\tilde{F}_{1}\sin ^{2}\theta \right)
\cos ^{2}\theta +\tilde{F}_{2}\sin ^{2}\theta \right] \sin ^{\left| n\right|
-1}\theta   \nonumber \\
H_{3} &=&-F_{3}\sin \theta +F_{4}\cos \theta   \nonumber \\
\  &=&\left[ \left( -f\left( r\right) -\tilde{F}_{3}\sin ^{2}\theta \right)
\sin ^{\left| n\right| -1}\theta +\tilde{F}_{4}\right] \sin \theta \cos
\theta   \nonumber \\
(1-H_{4}) &=&F_{3}\cos \theta +F_{4}\sin \theta   \nonumber \\
&=&\left[ f\left( r\right) +\tilde{F}_{3}\sin ^{2}\theta \right] \cos
^{2}\theta \sin ^{\left| n\right| -1}\theta +\tilde{F}_{4}\sin ^{2}\theta .
\label{10h}
\end{eqnarray}
This especially implies that $H_{1}$ and $1-H_{2}$ will always have a well
defined $\theta $-parity, whereas for $H_{3}$ and $1-H_{4}$ this is the case
only for odd $n$. At the point $r=0$ the gauge potential is analytical, if
moreover the functions $\tilde{F}_{i}$ have the from
\begin{eqnarray}
\tilde{F}_{i}\left( r,\sin ^{2}\theta \right)  &=&r^{2}\breve{F}_{i}\left(
r^{2},\sin ^{2}\theta \right) \text{ for }i=1,3  \nonumber \\
\tilde{F}_{i}\left( r,\sin ^{2}\theta \right)  &=&r\breve{F}_{i}\left(
r^{2},\sin ^{2}\theta \right) \text{ for }i=2,4
\end{eqnarray}
with analytical functions $\breve{F}_{i}$. The solutions constructed by
Kleihaus and Kunz \cite
{kleihaus1997a,kleihaus1997b,kleihaus1997c,kleihaus1998a,kleihaus1998b} do
not have the regular form (\ref{10h}) \cite{ft1}.

Kleihaus most recently showed \cite{kleihaus1999a} that for $n=2,3,4$ a
gauge transformation of the form 
\begin{equation}
U(\vec{r})=\exp (i\frac{\Gamma }{2}\tau _{\varphi }^{n})  \label{11}
\end{equation}
leads to a new gauge field 
\begin{eqnarray}
\hat{A} &=&UAU^{\dagger }+idUU^{\dagger }\   \nonumber \\
&=&\frac{1}{2gr}\left\{ 
\begin{array}{c}
\tau _{\phi }^{n}\left[ \hat{H}_{1}dr+(1-\hat{H}_{2})rd\theta \right]  \\ 
-n\left[ \hat{F}_{3}\tau _{\rho }^{n}+\hat{F}_{4}\tau _{3}\right] r\sin
\theta d\phi 
\end{array}
\right\}   \label{12}
\end{eqnarray}
that satisfies the following conditions: 
\begin{eqnarray}
\hat{H}_{1} &\sim &\rho ^{|n|}+O\left( \rho ^{|n|+1}\right)   \nonumber \\
1-\hat{H}_{2} &\sim &\rho ^{|n|+1}+O\left( \rho ^{|n|+2}\right)   \nonumber
\\
\hat{F}_{3} &=&[\frac{\rho }{r}\hat{H}_{3}+\frac{z}{r}(1-\hat{H}_{4})]\sim
\rho ^{|n|+1}+O\left( \rho ^{|n|+2}\right)   \nonumber \\
\hat{F}_{4} &=&[\frac{z}{r}\hat{H}_{3}-\frac{\rho }{r}(1-\hat{H}_{4})]\sim
\rho +O\left( \rho ^{2}\right)   \label{13}
\end{eqnarray}
We see that this transformation resulted in a continuous gauge potential 
\cite{ft2}, but since the conditions on the $\theta $-parity have not been
considered, the potentials are still possibly not differentiable at $\theta
=0$. One must at least show that the next to leading order vanishes, then
the potentials will be twice differentiable at least. Especially for even $n$
the $\theta $-parity of the functions $H_{1}$ and $H_{2}$, which is
necessary in view of (\ref{9}) and (\ref{10aa}) and (\ref{10ab}), is not
respected in the expansions given. Thus the desired aim, namely to show that
one can obtain all gauge invariant quantities also from a regular gauge
potential, that exists at least locally, has been failed. Moreover, where
expansions of the gauge potential functions $H_{i}$ have been carried
without respecting the necessary $\theta $-parity, we must indeed ask
whether regular solutions with axial symmetry exist at all. A proof of
existence is certainly needed, especially since more boundary conditions
must be satisfied than can be given to the numerical routines for solving
the nonlinear partial differential equations for these functions. Therefore,
at least so far as the $SU(2)$ Einstein-Yang-Mills(-dilaton) theory has been
studied by Kleihaus and Kunz, the numerical solutions give insufficient
indication for the existence of corresponding exact solutions.

Finally, a remark about the interpretation of the theory appears to be
indicated. Provided the existence of regular solutions has been established,
their discussion should not be based on gauge-dependent properties of some
singular potential. We have seen that the function $f$, which is the only
relevant one in the spherically symmetric case, has no physical meaning in
the axially symmetric case \cite{ft2}. For $n>1$ the functions $H_{2}$ and $%
H_{4}$ must satisfy $H_{2}=H_{4}=1$ on the symmetry axis for any regular
gauge potential, which means that the node number of these functions
introduced in the discussion are artifacts of the singular gauge. Hence the
classification of solutions that are not equivalent under gauge
transformations must be based on some other aspect.


\begin{references}
\bibitem{ft2}  The function $f$ can be removed by a gauge transformation
with $\Gamma =-f(r)\sin ^{\left| n\right| }\theta $.
\end{references}
\end{document}